\documentstyle[11pt]{article}

\columnwidth\textwidth

\def\kon#1#2{\vbox{\halign{##&&##\cr
\lower4pt\hbox{$\scriptscriptstyle\vert$}\hrulefill &
\hrulefill\lower4pt\hbox{$\scriptscriptstyle\vert$}\cr $#1$&
$#2$\cr}}}

\title{CAUSAL APPROACH TO SUPERSYMMETRY: CHIRAL SUPERFIELDS}
\author{Florin Constantinescu 
\\Fachbereich Mathematik,
\\ Johann Wolfgang Goethe Universit\"at Frankfurt,
\\ Robert-Mayer-Str. 10, D-60054 Frankfurt am Main, Germany
\and G\"unter Scharf 
\\ Institut f\"ur Theoretische Physik, 
\\ Universit\"at Z\"urich, 
\\ Winterthurerstr. 190 , CH-8057 Z\"urich, Switzerland}

\date{}
\begin{document}
\maketitle
\vskip 3cm
\begin{abstract}  
We construct quantized free superfields and
represent them as operator-valued distributions in Fock space
starting with Majorana fields.
The perturbative construction of the S-matrix for interacting theories
is carried through by extending the causal method of Epstein and Glaser
to superspace. We propose a scaling and singular order of distributions
in superspace by a procedure which scales both commutative and
non-commutative variables. Using this singular order the chiral
(Wess-Zumino) model appears to be super-renormalizable.

\end{abstract}
\newpage

\def\ro{\varrho}
\def\eh{{\scriptstyle{1\over 2}}}
\def\d{\partial}\def\=d{\,{\buildrel\rm def\over =}\,}
\def\dh{\mathop{\vphantom{\odot}\hbox{$\partial$}}}
\def\dl{\dh^\leftrightarrow}
\def\sqr#1#2{{\vcenter{\vbox{\hrule height.#2pt\hbox{\vrule width.#2pt 
height#1pt \kern#1pt \vrule width.#2pt}\hrule height.#2pt}}}}
\def\w{\mathchoice\sqr45\sqr45\sqr{2.1}3\sqr{1.5}3\,} 
\def\eps{\varepsilon}
\def\oe{\overline{\rm e}}
\def\onu{\overline{\nu}}
\def\ds{\hbox{\rlap/$\partial$}}
\def\psq{{\overline{\psi}}}
\def\la{{(\lambda)}}
\def\lap{\bigtriangleup\,}
\def\ev{{\scriptstyle{1\over 4}}}
\def\fii{\varphi}

\section{Introduction}

In the presentation of supersymmetry one mostly starts by representing
the supersymmetric algebra on {\it classical} fields [1-3]. Quantum
fields only appear at a later stage when propagators and Feynman rules
are derived. One exception is the third volume of Weinberg's book [4]
where some quantum fields are directly constructed without introducing
classical fields. In any case one is surprisingly brief about quantum
superfields. Some work concerning the axiomatic supersymmetric quantum
field theory is contained in [5] and [6]. We have looked in vain for 
commutation relations or a concrete Fock space representation of the 
free fields. If one does not want to rely on formal path-integral methods,
this is the essential basis of perturbation theory. Furthermore, the
complete understanding of the free quantum fields is indispensable for
the construction of gauge theories in the spirit of a recent monograph
[7]. For this reason we discuss the quantized Majorana
field in the next section which gives the fermionic component of the
superfield. The quantized chiral superfield is introduced in sect.3 by
applying finite susy-transformations to the scalar component. The
(anti-) commutation relations of the component fields yield the
commutation relations for the chiral superfield. Since the component
fields have their standard representations in the bosonic or fermionic
sectors of Fock space, the whole superfield is represented there as
well. A direct representation of the superfield without using component
fields is also possible, we will consider this subject elsewhere.

The construction of the free asymptotic superfield $\Phi$ enables us to
set up perturbation theory for the S-matrix, if a first order coupling
$T_1[\Phi]$ is given. For this purpose we use the causal method of
Epstein and Glaser [8] which we have to extend to superspace. We observe
that the causal structure in $x$-space can be established as in the case
of ordinary field theory [7, 9]. The subtle step in the construction of
time-ordered products is the splitting of causal distributions
$d(x,\theta,\bar\theta)$ into advanced and retarded parts. The latter is
obtained (in the so-called regular case) as the following distributional
limit
$$r(x,\theta,\bar\theta)=\lim_{\lambda\to 0}\chi_0\Bigl({v\cdot
x\over\lambda}\Bigl)d(x,\theta,\bar\theta),\eqno(1.1)$$
where $\chi_0(t)$ is a $C^\infty$-function on ${\bf R^1}$ which switches
from 0 to 1 in the intervall $0<t<1$ and $v$ is a fixed time-like vector. The
existence of the limit (1.1) depends on the behaviour of $d$ under
scaling transformation $x'=\lambda x$ for $\lambda\to 0$. However, the
scaling limit of $d$ only exists if the Grassmann variables
$\theta,\bar\theta$ are also scaled according to
$$\theta'=\sqrt{\lambda}\theta,\quad\bar\theta'=\sqrt{\lambda}\bar\theta.  
\eqno(1.2)$$
This leads us to modified definitions of quasi-asymptotics and singular
order $\omega$ on superspace in sect.4. The resulting formula for
$\omega$ shows that for the Wess-Zumino model only the second order
vacuum polarisation graphs allows renormalisation, apart from the vacuum
graph. This is the non-renormalisation theorem in the framework of the
causal theory.

\section{The quantized Majorana field}

We start from the quantized Dirac field 
$$\psi(x)=(2\pi)^{-3/2}\int d^3p\Bigl[b_s(\vec p)u_s(\vec p)e^{-ipx}+
d_s^+(\vec p)v_s(\vec p)e^{ipx}\Bigl]\eqno(2.1)$$
$$i\gamma^\mu\d_\mu\psi(x)=m\psi(x)\eqno(2.2)$$
in the chiral representation
$$\gamma^0=\pmatrix{0_2& 1_2\cr 1_2& 0_2\cr},\quad
\gamma^k=\pmatrix{0&-\sigma_k\cr \sigma_k&0\cr},\eqno(2.3)$$
where $\sigma_k$ are the Pauli matrices. The $u$- and $v$-spinors are
given by
$$u_s(\vec p)={1\over 2\sqrt{E(E+m)}}\pmatrix{(m+\sigma_\mu p^\mu)
\chi_s\cr (m+\hat\sigma_\mu p^\mu)\chi_s\cr}\eqno(2.4)$$
$$v_s(\vec p)={1\over 2\sqrt{E(E+m)}}\pmatrix{(m+\sigma_\mu p^\mu)
\chi_s\cr -(m+\hat\sigma_\mu p^\mu)\chi_s\cr}.\eqno(2.5)$$
Here
$$\sigma_\mu p^\mu=E+\vec\sigma\cdot\vec p=\sigma p,\quad
\hat\sigma_\mu p^\mu=E-\vec\sigma\cdot\vec p=\hat\sigma p,\eqno(2.6)$$
where $E=\sqrt{\vec p^2+m^2}$ and $\chi_s$ is the spin basis
$$\chi_1=\pmatrix{1\cr 0\cr},\quad\chi_{-1}=\pmatrix{0\cr 1\cr}.
\eqno(2.7)$$

The charge-conjugate spinor field is defined by
$$\psi_C=C\gamma^0\psi^{+T}=\pmatrix{-i\sigma_2& 0\cr 0& i\sigma_2\cr}
\pmatrix{0& 1\cr 1& 0\cr}\psi^{+T}=\pmatrix{0& -i\sigma_2\cr i\sigma_2
& 0\cr}\psi^{+T},\eqno(2.8)$$
where $C$ is the charge-conjugation matrix which can be expressed by
$\sigma_2$ in the chiral representation and the transposition $T$ refers
always to the spinor components, only. The Majorana field $\psi^M$ is 
defined by the property that it agrees with its charge-conjugate
$$\psi^M_C(x)=\psi^M(x).\eqno(2.9)$$
This implies the identification
$$d_s(\vec p)=-sb_{-s}(\vec p)\eqno(2.10)$$
in (2.1). Hence, the Majorana field is given by
$$\psi^M(x)=(2\pi)^{-3/2}\int d^3p\Bigl[u_s(\vec p)e^{-ipx}b_s(\vec p)-
sv_s(\vec p)e^{ipx}b_{-s}^+(\vec p)\Bigl],\eqno(2.11)$$
where the emission and absorption operators obey the usual
anti-commutation relations
$$\{b_s(\vec p), b_{s'}^+(\vec q)\}=\delta_{ss'}\delta^3(\vec p-
\vec q),\eqno(2.12)$$
and all other anti-commutators vanish. The fermionic Fock space is then
constructed in the usual manner by applying $b^+$'s to the vacuum
$\Omega$.

We are now in the position to calculate the anti-commutator between the
positive and negative frequency parts in (2.11)
$$\{\psi_M^{(-)}(x)_a,\,\psi_M^{(+)}(y)_b\}=
(2\pi)^{-3}\int d^3p\,u_{sa}(\vec p)sv_{-sb}(\vec p)^Te^{-ip(x-y)}.
\eqno(2.13)$$
To evaluate this we regard the r.h.s. as a $4\times 4$ matrix,
$a,b=1,\ldots 4$.
The matrix multiplication of the spin basis (2.7) gives the following
$4\times 4$ matrix
$$\pmatrix{\chi_s\cr\chi_s\cr}(\chi_{-s},\chi_{-s})s=\pmatrix{0& 1&
0& 1\cr -1& 0& -1& 0\cr 0& 1& 0& 1\cr -1& 0& -1& 0\cr}=\pmatrix{-\eps&
-\eps\cr -\eps&-\eps\cr}\eqno(2.14)$$
where the antisymmetric matrix
$$\eps_{ab}=\pmatrix{0& -1\cr 1& 0\cr}\eqno(2.15)$$
was introduced. Using
$$\hat\sigma_\mu=\eps\sigma_\mu^T\eps^{-1}\eqno(2.16)$$
and multiplying the various $2\times 2$ matrices we finally get for the
anti-commutator (2.13)
$$=(2\pi)^{-3}\int {d^3p\over 2E}\,e^{-ip(x-y)}\pmatrix{-m\eps& \sigma p
\eps\cr -\hat\sigma p\eps& m\eps\cr}.\eqno(2.17)$$
Without the matrix this would be equal to the positive frequency part of
the Jordan-Pauli distribution $-iD^{(+)}(x-y)$. The negative frequency
part is treated in the same way, yielding the following total
anti-commutator
$$\{\psi_M(x)_a,\,\psi_M(y)_b\}=\pmatrix{im\eps&
\sigma\d_x\eps\cr -\hat\sigma\d_x\eps& -im\eps\cr}D(x-y).\eqno(2.18)$$
Since $\hat\sigma\eps=\eps\sigma^T=-(\sigma\eps)^T$, the r.h.s. shows
the correct invariance under transposition and $x\leftrightarrow y$.

For supersymmetry one usually goes over to 2-component Weyl spinors in
the notation of van der Waerden. Concerning notation we mostly follow
the criterion of laziness: we prefer that notation which requires less
keys in TEX on the computer. Instead of dotted greek indices we
therefore use latin ones with a bar.
The Majorana field is then written as
$$\psi_M=\pmatrix{\chi_a\cr\bar\chi^{\bar b}\cr}.\eqno(2.19)$$
The lower spinor is obtained from the upper one according to (2.8-9)
$$\bar\chi^{\bar b}=-i\hat\sigma_2^{\bar ba}\chi_a^{+T}.\eqno(2.20)$$
Here we have written $-\hat\sigma_2$ instead of $\sigma_2$ in order to have
the indices at the correct places. It is essential to write the
antisymmetric matrix in (2.20) with $\hat\sigma_2$ because the Pauli
matrices have mixed indices in contrast to the $\eps$-tensor (2.15).
Lowering the index $\bar b$ by
multiplying with $\eps_{\bar a\bar b}$ and multiplying the $2\times 2$
matrices on the r.h.s., we find the simple result
$$\bar\chi_{\bar a}=(\chi^{+T})_a=\chi^+_a.\eqno(2.21)$$
We have omitted the transposition because it has no consequence if we
consider one spinor component only. It is not a definition but a
derived fact that charge conjugation of the quantized Majorana field 
is represented in Fock space simply by hermitian conjugation.

From (2.18) we can now read off the various anti-commutators
$$\{\chi_a(x),\chi_b(y)\}=im\eps_{ab}D(x-y)\eqno(2.22)$$
$$\{\chi_a(x),\chi^b(y)\}=-im\delta_{a}^bD(x-y)$$
$$\{\chi_a(x),\bar\chi^{\bar b}(y)\}=-\sigma_{a\bar c}^\mu\eps^{\bar c\bar
b}\d_\mu^x D(x-y)\eqno(2.23)$$
$$\{\chi_a(x),\bar\chi_{\bar b}(y)\}=\sigma^\mu_{a\bar b}\d_\mu^x 
D(x-y),\eqno(2.24)$$
$$\{\bar\chi^{\bar a}(x),\bar\chi^{\bar b}(y)\}=im\eps^{\bar a\bar b} 
D(x-y)\eqno(2.25)$$
$$\{\bar\chi_{\bar a}(x),\bar\chi_{\bar b}(y)\}=-im\eps_{\bar a\bar b} 
D(x-y)\eqno(2.26)$$
because the indices are raised and lowered with the anti-symmetric
$\eps$-tensor (2.15). The change of sign in (2.23) and (2.25) is due to the
relation
$$\eps_{ab}=-\eps^{\bar a\bar b},$$
since in (2.18) all $\eps$-matrices are equal to the original one (2.15).
The relation (2.26) can also be derived from (2.22) by means of charge
conjugation (2.21).
The Lorentz index in $\sigma^\mu$ is written upstairs from now on for
notational reason.

The Dirac equation (2.2) for the Majorana field (2.19) reads
$$i\sigma^\mu_{a\bar b}\d_\mu\bar\chi^{\bar b}=m\chi_a,\quad
i\hat\sigma^{\mu\bar ab}\d_\mu\chi_b=m\bar\chi^{\bar a}.\eqno(2.27)$$
Note that $\sigma^\mu\d_\mu=\sigma_0\d_0-\vec\sigma\cdot\vec d$, 
$\hat\sigma^\mu\d_\mu=\sigma_0\d_0+\vec\sigma\cdot\vec d$. Using
$$\hat\sigma^{\mu\bar ab}=\eps^{\bar a\bar b}\eps^{ba}\sigma^\mu_{a\bar
b}\eqno(2.28)$$ 
the second equation can also be written as
$$-i\sigma^\mu_{b\bar a}\d_\mu\chi^b=m\bar\chi_{\bar a}.
\eqno(2.29)$$

To get an explicit representation of the Mojorana field $\chi(f)$ in
Fock space we must smeare the first two components in (2.11) with
2-component test functions $f(x)\in S({\bf R^4})\otimes {\bf C^2}=E$. The
Majorana field is an operator in the antisymmetric Fock space ${\cal F}$
over ${\cal F}^1=E$. Using the well-known representation
$$\Bigl(b_s^+(\vec p)\fii\Bigl)_n={1\over\sqrt{n}}\sum_{j=1}^n(-1)^
{j-1}\delta(\vec p-\vec p_j)\chi_s\otimes\fii_{n-1}(p_1,s_1\ldots\hat p_j,
\hat s_j\ldots p_n,s_n)\eqno(2.30)$$
$$\Bigl(b_s(\vec p)\fii\Bigl)_n=\sqrt{n+1}\,\fii_{n+1}(p,s,p_1,s_1\ldots
p_n,s_n),\quad s_j=\pm 1,\eqno(2.31)$$
of the emission and absorption operators we find by means of (2.4-5)
$$\Bigl(\chi(f)\fii\Bigl)_n=\sqrt{2\pi}\biggl[\sqrt{n+1}\int{d^3p\over
2\sqrt{E(E+m)}}\hat f^T(-p)(m+\sigma p)\fii_{n+1}(p,p_1,\ldots p_n)$$
$$+{1\over\sqrt{n}}\sum_{j=1}^n(-1)^j{(m+\hat\sigma p_j)\eps\over
2\sqrt{E_j(E_j+m)}}\hat f(p_j)\otimes\fii_{n-1}(p_1,\ldots\hat p_j,
\ldots p_n)\biggl],\eqno(2.32)$$
$$\Bigl(\bar\chi(f)\fii\Bigl)_n=\sqrt{2\pi}\biggl[\sqrt{n+1}\int{d^3p\over
2\sqrt{E(E+m)}}\hat f^T(-p)(m+\hat\sigma p)\fii_{n+1}(p,p_1,\ldots p_n)$$
$$-{1\over\sqrt{n}}\sum_{j=1}^n(-1)^j{(m+\sigma p_j)\eps\over
2\sqrt{E_j(E_j+m)}}\hat f(p_j)\otimes\fii_{n-1}(p_1,\ldots\hat p_j,
\ldots p_n)\biggl].\eqno(2.33)$$
The argument $\hat p_j$ has to be left out.

\section{Quantized chiral superfield}

We want to represent the supersymmetric algebra
$$\{Q_a,\bar Q_{\bar b}\}=2\sigma^\mu_{a\bar b}P_\mu=-2i
\sigma^\mu_{a\bar b}\d_\mu,\eqno(3.1)$$
$$\{Q_a, Q_b\}=0=\{\bar Q_a,\bar Q_{\bar b}\}=[Q_a, P_\mu]\eqno(3.2)$$
by operators in Fock space. $Q_a$ and $\bar Q_{\bar b}$ transform
according to the $(\eh, 0)$ and $(0, \eh)$ representations of the proper
Lorentz group, respectively. As usual one rewrites (3.1) as a
Lie-algebra commutator
$$[\xi^a Q_a,\bar\xi^{\bar b}\bar Q_{\bar b}]=2\xi\sigma\bar\xi P\eqno(3.3)$$
by introducing anti-commuting C-numbers $\xi^a, \bar\xi^{\bar b}$.
Most authors discuss classical superfields first, we exclusively work
with free quantized fields in Fock space. So it is our aim to construct 
a field $\Phi$ with the transformation law
$$\delta_\xi\Phi=[\xi Q+\bar\xi\bar Q,\Phi]\eqno(3.4)$$
under infinitesimal supersymmetric transformations. Taking as initial
value a scalar field $A(x)$, the solution is given by the finite
transformation
$$\Phi(x,\theta,\bar\theta)=e^{\theta Q+\bar\theta\bar Q}A(x)
e^{-\theta Q-\bar\theta\bar Q}.\eqno(3.5)$$
We assume $A(x)$ to be a charged scalar field of mass $m$, quantized in the
usual way
$$[A(x), A^+(x')]=-iD(x-x'),\eqno(3.6)$$
where $D(x)$ is the causal Jordan-Pauli distribution and all other
commutators vanish.

We are now going to calculate (3.5) by means of Hausdorff's formula
$$\Phi(x,\theta,\bar\theta)=e^{\theta Q}e^{\bar\theta\bar Q}
e^{-\theta\sigma\bar\theta P}A(x)e^{\theta\sigma\bar\theta P}
e^{-\bar\theta\bar Q}e^{-\theta Q},\eqno(3.7)$$
where (3.3) has been used. By the Lie series we first compute
$$A'\=d e^{-\theta\sigma\bar\theta P}A(x)e^{\theta\sigma\bar\theta P}=$$
$$=A+[i\theta\sigma\bar\theta\d, A]+\eh [i\theta\sigma\bar\theta\d,
[i\theta\sigma\bar\theta\d,A]],\eqno(3.8)$$
using $P_\mu=-i\d_\mu$.
The first commutator gives $i\theta\sigma^\mu\bar\theta(\d_\mu A)$ and
then the double commutator becomes equal to
$$-\eh\theta\sigma^\mu\bar\theta\theta\sigma^\nu\bar\theta[\d_\mu,
(\d_\nu A)]=-\ev\theta\theta\bar\theta\bar\theta\w A,
\eqno(3.9)$$
here we have used the identity
$$\theta\sigma^\mu\bar\theta\theta\sigma^\nu\bar\theta=\eh
\eta^{\mu\nu}\theta\theta\bar\theta\bar\theta.\eqno(3.10)$$
The result can also be written as follows
$$A'=\Bigl(e^{i\theta\sigma\bar\theta\d}A(x)\Bigl)\=d A(x+i
\theta\sigma\bar\theta)\=d A(y),$$
$$A(y)=A(x)+i\theta\sigma\bar\theta\d A(x)-\ev\theta\theta\bar\theta\bar
\theta\w A(x),\eqno(3.11)$$
with the natural variable
$$y=x+i\theta\sigma\bar\theta\eqno(3.12)$$
in super-space.

To continue the calculation of (3.7) we need the commutators of $\bar 
Q_{\bar a}, Q_a$ with $A(x)$. Let us assume
$$[\bar Q_{\bar a},A(x)]=0,\eqno(3.13)$$
this defines the chiral superfield in contrast to the anti-chiral one,
and
$$[Q_a, A(x)]=\sqrt{2}\psi_a(x).\eqno(3.14)$$
Like $Q_a$ $\psi$ must be a 2-component spinor field transforming by the $(\eh,
0)$-representation, the factor $\sqrt{2}$ is conventional. 
Comparing (3.1) with (2.22-26) we see that the anti-commutation rules of the
$Q$'s are the same as for (massless) Majorana fields. Therefore, we quantize
the spinor field $\psi$ as a Majorana field with mass $m$:
$$\{\psi_a(x),\psi_b(x')\}=im\eps_{ab}D(x-x')\eqno(3.15)$$
$$\{\bar\psi^{\bar a}(x),\bar\psi^{\bar b}(x')\}=im\eps^{\bar a\bar
b}D(x-x') \eqno(3.16)$$
$$\{\psi_a(x),\bar\psi_{\bar b}(x')\}=\sigma^\mu_{a\bar b} 
\d_\mu^x D(x-x').\eqno(3.17)$$
The rule (3.14) can also be used for $A(y)$ (3.11), hence
$$e^{\theta Q}A(y)e^{-\theta Q}=A(y)+\sqrt{2}\theta\psi(y)+\eh
[\theta Q,[\theta Q,A]].\eqno(3.18)$$

The double commutator is equal to
$$\eh\sqrt{2}[\theta^a Q_a,\theta^b\psi_b(y)]=-{1\over\sqrt{2}}\theta^a
\theta^b\{Q_a,\psi_b(y)\}.\eqno(3.19)$$
The anti-commutator on the r.h.s. must be anti-symmetric in $a, b$, it
can thus be written as
$$\{Q_a,\psi_b(y)\}=\sqrt{2}\eps_{ab}F(y),\eqno(3.20)$$
where $F(y)$ is a scalar field. Then we arrive at the usual expression
$$\Phi=A(y)+\sqrt{2}\theta\psi(y)+\theta\theta F(y).\eqno(3.21)$$
If we use the expansion (3.11) for $A(y)$ and $\psi(y)$ we obtain
$$\Phi(x,\theta,\bar\theta)=A(x)+i\theta\sigma\bar\theta\d A 
-\ev\theta\theta\bar\theta\bar\theta\w A(x)+$$
$$+\sqrt{2}\theta^a(\psi_a(x)+i\theta\sigma\bar\theta\d\psi_a)+
\theta\theta F(x).\eqno(3.22)$$
Using the identity
$$\theta^a\theta^b=-\eh\eps^{ab}\theta\theta,\eqno(3.23)$$
the second term in the bracket is equal to
$$-{i\over\sqrt{2}}\theta\theta\d\psi\sigma\bar\theta.\eqno(3.23)$$

The scalar field $F(x)$ is not independent. Indeed, from (3.15) and
(3.14) we find by means of the Jacobi identity
$$\{\psi_a(x),\psi_b(x')\}={1\over\sqrt{2}}\{\psi_a(x),[Q_b,A(x')]\}=$$
$$={1\over\sqrt{2}}\Bigl(\{Q_b,[A(x'),\psi_a(x)]\}-[A(x'),\{
\psi_a(x),Q_b\}]\Bigl)=$$
$$=-\eps_{ab}[F(x), A(x')]=im\eps_{ab}D(x-x').\eqno(3.24)$$
Since the adjoint of (3.13) implies
$$[F(x), A^+(x')]=0$$
in a similar way, we may conclude
$$F(x)=-mA^+(x),\eqno(3.25)$$
up to a C-number field which we set equal to 0. All component fields,
i.e. $A(x)$ and $\psi(x)$, have their natural representations as
operator-valued distributions in a big Fock space, hence, the chiral
superfield is represented as well. The $\theta, \bar\theta$ are merely
spectators in this construction. Since we know all commutators of the
supercharges $Q, \bar Q$ with the fundamental fields $A, \psi$, we are
also able to find the representation of the supercharges, but this is
not needed for the following.

From the commutation rules of the component fields it is now easy to
determine the commutator of the chiral superfield. This is best done
with the expression (3.21). Taking (3.25) into account we find
$$[\Phi(y,\theta),\Phi(y',\theta')]=\theta'\theta'[A(y),-mA^+(y')]+
\theta\theta[-mA^+(y),A(y')]+$$
$$+2\theta^a\theta^{\prime b}\{\psi_a(y),\psi_b(y')\}=im(\theta-\theta')^2
D(y-y').\eqno(3.26)$$
The square can also be regarded as the 2-dimensional
$\delta$-distribution
$$(\theta-\theta')^2=\delta^2(\theta-\theta')\eqno(3.27)$$
because its integral $\int d^2(\theta -\theta')$ is one. By the translation
formula
$$D(y-y')=D(x+i\theta\sigma\bar\theta-x'-i\theta'\sigma\bar\theta')=$$
$$=\Bigl[\exp i(\theta\sigma\bar\theta-\theta'\sigma\bar\theta')\d_x
\Bigl]D(x-x')\eqno(3.28)$$
we can go over to ordinary variables in superspace:
$$[\Phi(x,\theta,\bar\theta),\Phi(x',\theta',\bar\theta')]=im\delta^2
(\theta-\theta')\Bigl[\exp i(\theta\sigma\bar\theta-\theta'\sigma\bar\theta')\d
_x\Bigl]D(x-x').\eqno(3.29)$$

The adjoint of the chiral superfield is the anti-chiral one. To see this
one has to use adjoining relations like
$$(\theta\psi)^+=\bar\theta\bar\psi\eqno(3.30)$$
which follow from (2.20). In the same way as above one obtains the
following further commutation relations
$$[\Phi(x,\theta,\bar\theta),\Phi^+(x',\theta',\bar\theta')]=-i
\Bigl[\exp i(\theta\sigma\bar\theta+\theta'\sigma\bar\theta'-2 
\theta\sigma\bar\theta')\d_x\Bigl]D(x-x')\eqno(3.31)$$
$$[\Phi^+(x,\theta,\bar\theta),\Phi^+(x',\theta',\bar\theta')]=-im\delta^2
(\bar\theta-\bar\theta')\Bigl[\exp-i(\theta\sigma\bar\theta- 
\theta'\sigma\bar\theta')\d_x\Bigl]D(x-x').\eqno(3.32)$$
The commutators of the various emission and absorption parts are of the
same form, only $D$ must be substituted by $D^{(+)}$ or $D^{(-)}$.
If one expands the exponentials one gets up to two derivatives on the
Jordan-Pauli distribution $D(x-x')$. Considered as distributions in
$x$-space alone, these are rather singular distributions and it is not
at all clear why supersymmetric theories have a tame ultraviolet
behaviour. The point is that we have to consider all distributions in
superspace. Then, as we will see in the next section, the distributions
in the commutation relations are as good as in ordinary field theory. 

\section{Causal perturbation theory}

Our goal is to construct the S-functional $S(g)$ as a formal power
series
$$S(g)=\sum_{n=0}^\infty {1\over n!}\int dX_1\ldots dX_n\,T_n(X_1,\ldots
,X_n)g(X_1)\ldots g(X_n).\eqno(4.1)$$
Here $X_j=(x_j^\mu,\theta_j,\bar\theta_j)$ denotes the superspace
coordinates, $dX_j=d^4x_jd^2\theta_jd^2\bar\theta_j$, and $g(X_j)$ is a
test-function on superspace, the physical S-matrix is obtained in the
adiabatic limit $g\to 1$. The time-ordered products $T_n$ are
constructed inductively, $T_1(X)$ must be given. For the Wess-Zumino
model it reads
$$T_1(X)=i\bigl({\lambda^3\over 3!}:\Phi(X)^3:\delta^2(\bar\theta) 
+{\rm h.c.}\bigl).\eqno(4.2)$$
The double dots denote normal ordering of the free superfields.

The essential property of $T_1(X)$ is the causal commutator in $x$-space
$$[T_1(X_1), T_1(X_2)]=0,\quad (x_1-x_2)^2<0.\eqno(4.3)$$
This follows from the causal support of $D(x)$ in the commutation
relations. In $n$-th order we require the following causal factorization
$$T_n(X_1,\ldots X_n)=T_m(X_1,\ldots X_m)T_{n-m}(X_{m+1},\ldots X_n)
\eqno(4.4)$$
for $\{x_1,\ldots x_m\}>\{x_{m+1},\ldots x_n\}$ and arbitrary
$\theta$'s. The last inequality means
$$\{x_1,\ldots x_m\}\subset\Bigl\{\{x_{m+1},\ldots x_n\}+\bar V_-\Bigl
\}=0,\eqno(4.5)$$
where $\bar V_-$ is the closed backward light-cone. In addition the
$T_n$'s are assumed to be symmetric in all arguments and translation
invariant. 

The above properties allow the inductive construction of the $T_n$'s by
the method of Epstein and Glaser. This construction goes as follows:
Assuming $T_m(X_1,\ldots X_m)$ for $m\le n-1$ to be known, then we can
compute
$$A'_n(X_1,\ldots X_n)=\sum_{P_2}\tilde T_{n_1}(X)T_{n-n_1}(Y,X_n)
\eqno(4.6)$$
$$R'_n(X_1,\ldots X_n)=\sum_{P_2}\tilde T_{n-n_1}(Y,X_n)\tilde T_{n_1}
(X),\eqno(4.7)$$
where the sums run over all partitions
$$P_2:\quad\{X_1,\ldots X_n\}=X\cup Y,\quad X\ne\emptyset\eqno(4.8)$$
into disjoint subsets with $|X|=n_1\ge 1$, $|Y|\le n-2$. 
The $\tilde T_n$ are obtained by inversion of the perturbation series
$S(g)^{-1}$, they can be expressed by the $T_n$'s as follows
$$\tilde T_n(X)=\sum_{r=1}^n(-)^r\sum_{P_r}T_{n_1}(X_1)\ldots
T_{n_r}(X_r),$$
where the second sum runs over all partitions $P_r$ of $X$ into $r$
disjoint subsets. 

If the sums in (4.6-7) are
extended over all partitions $P_2^0$, including the empty set
$X=\emptyset$, then we get the distributions
$$A_n(X_1,\ldots X_n)=\sum_{P_2^0}\tilde T_{n_1}(X)T_{n-n_1}(Y,X_n)$$
$$=A'_n+T_n(X_1,\ldots X_n),\eqno(4.8)$$
$$R_n(X_1,\ldots X_n)=\sum_{P_2^0}\tilde T_{n-n_1}(Y,X_n)\tilde T_{n_1}(X)$$
$$=R'_n+T_n(X_1,\ldots X_n).\eqno(4.9)$$
These two distributions are not known by the induction assumption
because they contain the unknown $T_n$. Only the difference
$$D_n=R'_n-A'_n=R_n-A_n\eqno(4.10)$$
is known. Here the causal structure comes in: $D_n$ has causal support,
$R_n$ is the retarded part and $-A_n$ the advanced part
$${\rm supp}\,R_n(X_1,\ldots X_n)\subseteq\Gamma^+_{n-1}(x_n)$$
$${\rm supp}\,A_n(X_1,\ldots X_n)\subseteq\Gamma^-_{n-1}(x_n), 
\eqno(4.11)$$
with
$$\Gamma^\pm_{n-1}(x_n)=\{(X_1,\ldots X_n)| x_j\in\bar V_\pm(x_n),
\forall j=1,\ldots n-1\},\eqno(4.12)$$
where
$$\bar V_-(x)=\{y|(y-x)^2\ge 0, y^0\le x^0\}$$
is the closed backward cone in Minkowski space and $\bar V_+$ the closed
forward cone. These support properties follow from causality (4.4) in 
the same way as for ordinary quantum field theory. What remains to be
done is the splitting of $D_n$ (4.10) into retarded and advanced parts,
then $T_n$ follows from (4.8-9)
$$T_n=R_n-R'_n=A_n-A'_n.\eqno(4.13)$$

The distribution splitting is achieved by expanding $D_n$ in normally
ordered form using Wick's theorem for superfields and splitting the
numerical distributions. To split a numerical distribution $d(X)$ on
superspace, it is necessary to analyse the behaviour in the
neighbourhood of $X=0$ by means of scaling transformations
$$x'_j=\lambda x_j,\quad\theta'=\sqrt{\lambda}\theta,\quad
\bar\theta'=\sqrt{\lambda}\bar\theta.\eqno(4.14)$$
The $\theta$'s must be scaled is this way to get non-trivial limits for
$\lambda\to 0$ (cf. (3.28)). As in ordinary field theory we introduce
the following definitions:

{\bf Definition 1.} The distribution $d(X)$, tempered in $m$-dimensional 
$x$-space, has a quasi-asymtotics $d_0(X)$ at $X=0$ with respect to a 
positive continuous function $\ro(\lambda), \lambda>0$, if the limit
$$\lim_{\lambda\to 0}\ro(\lambda)\lambda^m d(\lambda X)=d_0(X)
\ne 0\eqno(4.15)$$
exists.

There is an equivalent definition for the Fourier transform with respect
to $x$:

{\bf Definition 2.} The distribution $\hat d(p,\theta,\bar\theta)$ has
quasi-asymptotics $\hat d_0(p,\theta,\bar\theta)$ if the limit
$$\lim_{\lambda\to 0}\ro(\lambda)\hat d\Bigl({p\over\lambda},\sqrt{ 
\lambda}\theta,\sqrt{\lambda}\bar\theta\Bigl)=\hat d_0(p,\theta, 
\bar\theta)\ne 0\eqno(4.16)$$
exists. If the function $\ro(\lambda)$ satisfies
$$\lim_{\lambda\to 0}{\ro(a\lambda)\over\ro(\lambda)}=a^\omega
\eqno(4.17)$$
for each $a>0$, then $d$ is called singular of order $\omega$ at $X=0$.
$\ro(\lambda)$ is called power counting function.

It is easy to check that the distribution on the r.h.s. of the
commutation relation (3.31) has $\omega=-2$, as in ordinaty field theory.
The distributions in (3.29) and (3.32) have even $\omega=-3$ due to the
additional $\delta$-distributions. If a causal distribution $d(X)$ has 
negative $\omega$ it can be splitted trivially by multiplying with a
step-function $\Theta(x^0)$. This can again be proven as in ordinary
field theory. For $\omega\ge 0$ the splitting must be carefully done, in
the usual terminology the graph needs renormalization. The
non-renormalization property of the Wess-Zumino model with coupling
(4.2) in the causal approach is now contained in the following theorem:

{\bf Theorem:} The singular order of a super-graph of order $n$ for the 
Wess-Zumino model satisfies
$$\omega\le 4-{1\over 2}(e+e')+l_+-{5\over 2}n,\eqno(4.18)$$
where $e$ and $e'$ is the number of external $\Phi$- and $\Phi^+$-legs
and $l_+$ the number of $\Phi\Phi^+$-contractions.

{\bf Proof.} The proof is inductively. In the inductive step we must form
tensor products of two distributions
$$T_r^1(x_1,\theta_1,\bar\theta_1,\ldots x_r,\theta_r,\bar\theta_r)
T_s^2(y_1,\theta'_1,\bar\theta'_1,\ldots y_s,\theta'_s,\bar\theta'_s).
\eqno(4.19)$$
Let us assume that in the normal ordering $l$ $\Phi\Phi$- or
$\Phi^+\Phi^+$-contractions are performed. Using translation invariance,
the resulting numerical distributions are of the follwing form
$$t_1(x_1-x_r,\ldots x_{r-1}-x_r;\theta,\bar\theta)\prod_{j=1}^l D^{(+)}
(x_{r_j}-y_{s_j};\theta_{r_j},\bar\theta_{r_j},\theta'_{s_j},\bar 
\theta'_{s_j})\times$$
$$\times t_2(y_1-y_s,\ldots y_{s-1}-y_s;\theta',\bar\theta').\eqno(4.20)$$
We introduce relative coordinates
$$\xi_j=x_j-x_r,\ \eta_j=y_j-y_s,\ \eta=x_r-y_s,$$
and compute the Fourier transform in $x$-space. Since the products go
over into convolutions, we get
$$\hat t(p_1,\ldots p_{r-1},q_1,\ldots q_{s-1},q;\theta,\bar\theta,
\theta',\bar\theta')=\int t(\xi,\eta,\theta,\bar\theta,\theta',\bar
\theta')e^{ip\xi+iq\eta}d^{4r-4}\xi\,d^{4s}\eta$$
$$=\int\prod_{j=1}^ld\kappa_j\delta\Bigl(q-\sum_{j=1}^l\kappa_j\Bigl)
\hat t_1(\ldots p_i-\kappa_{r(i)}\ldots;\theta,\bar\theta)$$
$$\times\prod_{j=1}^l\hat D^{(+)}(\kappa_j;\theta_{r_j},\bar\theta_{r_j},
\theta'_{s_j},\bar\theta'_{s_j})\hat t_2
(\ldots q_i+\kappa_{s(i)}\ldots;\theta',\bar\theta').\eqno(4.21)$$

Applying this to a test function $\varphi$, we have
$$\langle\hat t,\varphi\rangle=\int d^{4r-r}p'\,d^{4s-4}q'\,d^{4r}\theta
\,d^{4s}\theta'\,\hat t_1(p',\theta,\bar\theta)\times$$
$$\times \hat t_2(q',\theta',\bar\theta')\psi(p',q',\theta,\bar\theta,
\theta',\bar\theta')\eqno(4.22)$$
with
$$\psi(\ldots)=\int d^4q\prod_jd\kappa_j\,\delta\Bigl(q-\sum_j\kappa_j
\Bigl)\varphi(\ldots p'_i+\kappa_{r(i)},\ldots q'_i-\kappa_{s(i)},
\ldots q;\theta,\bar\theta,\theta',\bar\theta')$$
$$\times\prod_{j=1}^l\hat D^{(+)}(\kappa_j;\theta_{r_j},\bar\theta_{r_j},
\theta'_{s_j},\bar\theta'_{s_j}).\eqno(4.23)$$

In order to determine the singular order of $\hat t$ we have to consider
the scaled distribution
$$\Bigl\langle\hat t\Bigl({p\over\lambda},\sqrt{\lambda}\theta,\ldots\Bigl),
\varphi\Bigl\rangle=\lambda^{4(r+s-1)}\lambda^{-2(r+s)}\Bigl\langle
\hat t(p),\varphi\Bigl(\lambda p,{\theta\over\sqrt{\lambda}},\ldots
\Bigl\rangle=$$
$$=\lambda^{2(r+s-2)}\int d^{4r-4}p'\, d^{4s-4}q'\,d^{4r}\theta\,d^{4s}
\theta'\,\hat t_1(p',\theta,\bar\theta)\times$$
$$\times\hat t_2(q',\theta',\bar\theta')\psi_\lambda(p',q',\theta,
\bar\theta,\theta',\bar\theta')\eqno(4.24)$$
where
$$\psi_\lambda=\int d^4q\prod_{j=1}^ld\kappa_j\,\delta^1\Bigl(q-\sum_j
\kappa_j\Bigl)\varphi\Bigl(\ldots,\lambda(p'_i+\kappa_{r(i)}),\ldots
\lambda(q'_i-\kappa_{s(i)}),\lambda q;{\theta\over\sqrt{\lambda}},
{\bar\theta\over\sqrt{\lambda}},{\theta'\over\sqrt{\lambda}},
{\bar\theta'\over\sqrt{\lambda}}\Bigl)$$
$$\times\prod_{j=1}^l\hat D^{(+)}(\kappa_j;\theta_{r_j},\bar\theta_{r_j},
\theta'_{s_j},\bar\theta'_{s_j}).\eqno(4.25)$$
We introduce scaled variables $\tilde\kappa_j=\lambda\kappa_j$, $\tilde
q=\lambda q$, $\tilde\theta=\theta/\sqrt{\lambda}$ etc. and note that
$$\lim_{\lambda\to 0}\lambda^{-3}\hat D^{(+)}\Bigl({\kappa\over\lambda},
\tilde\theta\sqrt{\lambda},\ldots\Bigl)=\hat D_0^{(+)}(\tilde\kappa,
\tilde\theta,\ldots).$$
This implies for $\lambda\to 0$
$$\psi_\lambda\to{1\over\lambda^l}\psi\Bigl(\lambda p',\lambda q',
{\theta\over\sqrt{\lambda}},{\bar\theta\over\sqrt{\lambda}},
{\theta'\over\sqrt{\lambda}},{\bar\theta'\over\sqrt{\lambda}}\Bigl).
\eqno(4.26)$$
Using scaled variables $\lambda p'=\tilde p$,$\lambda q'=\tilde q$,
$\theta/\sqrt{\lambda}=\tilde\theta,\ldots$ again, we find
$$\Bigl\langle\hat t\Bigl({p\over\lambda},\sqrt{\lambda}\theta,\ldots\Bigl),
\varphi\Bigl\rangle\to{\lambda^4\over\lambda^l}
\int d^{4r-4}p'\, d^{4s-4}q'\,d^{4r}\theta\,d^{4s}
\theta'\,\hat t_1({\tilde p\over\lambda},\sqrt{\lambda}\theta,\sqrt
{\lambda}\bar\theta)\times$$
$$\times\hat t_2({\tilde q\over\lambda},\sqrt{\lambda}\theta',\sqrt
{\lambda}\bar\theta')\psi(\tilde p,\tilde q,\theta,
\bar\theta,\theta',\bar\theta').\eqno(4.27)$$

By the induction hypothesis, $\hat t_1$ and $\hat t_2$ have singular
order $\omega_1$, $\omega_2$ with power counting functions $\ro_1(
\lambda)$, $\ro_2(\lambda)$, respectively. Then the following limit
exists:
$$\lim_{\lambda\to 0}\lambda^{l-4}\ro_1(\lambda)\ro_2(\lambda)
\Bigl\langle\hat t\Bigl({p\over\lambda},\sqrt{\lambda}\theta,\ldots\Bigl),
\varphi\Bigl\rangle=\langle\hat t_0(p,\theta,\ldots),\varphi\rangle.
\eqno(4.28)$$
Hence, the singular order of $\hat t(p,\theta,\bar\theta)$ satisfies
$$\omega\le\omega_1+\omega_2+l-4.\eqno(4.29)$$
If instead of the $l$ $\Phi\Phi$-contractions we make $l_+$
$\Phi\Phi^+$-contractions, then in the same way we obtain
$$\omega\le\omega_1+\omega_2+2l_+-4.\eqno(4.30)$$

The results (4.29-30) suggest the following ansatz
$$\omega\le 4+l+l'+2l_++\alpha n.\eqno(4.31)$$
$l$ is the number of $\Phi\Phi$- $l'$ the number of $\Phi^+\Phi^+$-
and $l_+$ the $\Phi\Phi^+$-contractions, 4 is a consequence of
translation invariance. The term $\alpha n$ (with $\alpha$ still
unknown) is possible because the order of perturbation theory is
additive. Consider now a supergraph with $n_1$ $\Phi^3$-vertices and
$n_2$ $\Phi^{+3}$-vertices, $n_1+n_2=n$. Then for the number $e$ of
external $\Phi$-legs and $e'$ external $\Phi^+$-legs we have the
relations
$$3n_1=e+2l+l_+,\quad 3n_2=e'+2l'+l_+.\eqno(4.32)$$
Solving for $l$ and $l'$ and substituting into (4.31), we obtain
$$\omega\le 4+{3\over 2}(n_1+n_2)-{1\over 2}(e+e')-l_++2l_++\alpha n$$
$$=4-{1\over 2}(e+e')+l_++\Bigl({3\over 2}+\alpha\Bigl)n.\eqno(4.33)$$
The value of $\alpha$ follows from the beginning of the induction: for
$n=1$ $\omega$ must be zero. Taking one $\Phi^3$-vertex we have
$$\omega=4-{3\over 2}+{3\over 2}+\alpha=0,$$
which gives $\alpha=-4$. Hence
$$\omega\le 4-{1\over 2}(e+e')+l_+-{5\over 2}n,\eqno(4.34)$$
this completes the proof.

\section{Conclusions}

The minus sign $-5n/2$ in (4.34) shows that the theory is
super-renormalizable, because for $n$ big enough $\omega$ becomes
negative. This is obviously a consequence of the purely polynomial
$\Phi^3$-structure of the vertex. For an ordinary scalar $\varphi^3$-theory
which is super-normalizable as well one has
$$\omega\le 4-e-n.\eqno(5.1)$$
The bigger factor -5/2 in (4.34) shows that the Wess-Zumino model has
still a better ultraviolet behaviour. In fact, apart from the vacuum graph,
there is only one graph
with non-negative $\omega$: this is the second-order loop with one
$\Phi^3$- and one $\Phi^{+3}$-vertex which has
$$\omega=4-1+2-5=0.\eqno(5.2)$$
If one adds an additional inner line, $l_+$ increases by 1 but the order
$n$ by 2, consequently $\omega$ becomes more negative. Therefore, the
second-order loop graph (5.2) is the only one which requires
normalisation. This is the non-renormalisation property for time-ordered
products $T_n(X)$ on superspace.
The formula (4.34) gives an upper bound for the singular order. Indeed,
for certain graphs $\omega$ is actually smaller as a consequence of
supersymmetry. But this is not important because $\omega$ is anyway
negative so that no renormalisation is needed.
The singular order considered here, which is the essential parameter for
the distribution splitting, is not the same as the degree of
divergence $D$ considered by other authors [2, 3]. The latter refers in
our language to the $T_n$ after integration over $\theta$'s. We will 
return to this point elsewhere.

\end{document}